# A systematic approach for quantitative orientation and phase fraction analysis of thin films through grazing incidence X-ray diffraction

Fabian Gasser,[a] Sanjay John,[a] Jorid Smets,[b] Josef Simbrunner,[c] Mario Fratschko,[a] Víctor Rubio-Giménez,[b,d] Rob Ameloot,[b] Hans-Georg Steinrück[e,f] and Roland Resel[a]

[a] Institute of Solid State Physics, Graz University of Technology, Petersgasse 16, 8010 Graz, Austria

[b] Center for Membrane Separations, Adsorption, Catalysis and Spectroscopy (cMACS), KU Leuven, Celestijnenlaan 200f, 3001 Leuven, Belgium

[c] Department of Neuroradiology, Vascular and Interventional Radiology, Medical University Graz, Auenbruggerplatz 9, 8036 Graz, Austria

[d] Instituto de Ciencia Molecular (ICMol), Universitat de València, Catedrático José Beltrán 2, 46980 Paterna, Spain

[e] Institute for a Sustainable Hydrogen Economy (INW), Forschungszentrum Jülich GmbH, Marie-Curie-Straße 5, 52428 Jülich, Germany

[f] Institute of Physical Chemistry, RWTH Aachen University, Landoltweg 2, 52074 Aachen, Germany

**Synopsis**  This work introduces an algorithm to extract quantitative orientation and phase information of thin films using grazing incidence X-ray diffraction. The approach is demonstrated using experimental data from three distinct systems, showcasing its broad applicability.

**Abstract**  Grazing incidence X-ray diffraction (GIXD) is widely used for the structural characterization of thin films, particularly for analyzing phase composition and the orientation distribution of crystallites. While various tools exist for qualitative evaluation, a widely applicable systematic procedure to achieve quantitative information has not yet been developed. This work presents a first step in that direction, allowing accurate quantitative information to be achieved through the evaluation of radial line profiles from GIXD data. An algorithm is introduced for computing radial line profiles based on the crystal structure of known compounds. By fitting experimental data with calculated line profiles, accurate quantitative information about orientation distribution and phase composition is obtained, along with additional parameters such as mosaicity and total crystal volume. The approach is demonstrated using three distinct thin film systems, highlighting the broad applicability of the algorithm. This method provides a systematic and general approach to obtaining quantitative information from GIXD data.

## 1. Introduction

Thin films of organic molecules and hybrid inorganic-organic compounds promise a wide range of applications, including organic electronics (Köhler & Bässler, 2015; Kalyani *et al.*, 2017), optoelectronics (Mingabudinova *et al.*, 2016; Tian *et al.*, 2021), energy storage (Bai *et al.*, 2021; Li *et al.*, 2017), and chemical separation (Liu *et al.*, 2009; Guo *et al.*, 2009). The performance and stability of thin film devices are strongly related to their structural properties. Controlling parameters such as the degree of crystallinity (Rivnay *et al.*, 2012; Mahmood & Wang, 2020) and phase purity (Liang Tan *et al.*, 2020) is therefore essential. In the case of anisotropic crystals, properties vary with certain directions in space, leading to a strong demand to characterize and optimize the orientation distribution of crystallites, i.e. the texture, in polycrystalline films (Khalil *et al.*, 2023). A well-established tool for obtaining accurate crystallographic information from thin films is grazing incidence X-ray diffraction (GIXD) (Werzer *et al.*, 2024). GIXD does not require extensive sample preparation and is applicable in liquid, gaseous and vacuum environments. Area detectors allow to measure large volumes of reciprocal space in a single exposure (Schlepütz *et al.*, 2005), providing complete phase and texture information within short time frames. The combination of short measurement times and its non-destructive nature makes GIXD an excellent tool for both in situ (Greco *et al.*, 2018) and operando studies (Paulsen *et al.*, 2020), allowing real-time correlation between device performance and structural properties.

For many experiments, a qualitative analysis of GIXD data does not provide sufficient information. In particular, when different phases or complex orientation distributions are present in the same sample, a quantitative characterization is desirable (Ogle *et al.*, 2019; Steele *et al.*, 2023). For example, quantitative information is useful for optimizing the preparation conditions of functional thin films (Fischer *et al.*, 2023; Grott *et al.*, 2022; Müller-Buschbaum, 2014). Similarly, for in situ GIXD measurements during sample preparation quantitative analysis is of interest, as it can help to elucidate the film growth kinetics (Qin *et al.*, 2021; Reus *et al.*, 2022; Chou *et al.*, 2013; Richter *et al.*, 2015). Although such quantifications are widely used in literature, the theoretical background and the required intensity corrections have not been fully elaborated so far.

In this work, we introduce a general approach for accurate quantification of both orientation distribution and relative amount of phases within thin film samples by applying a single algorithm. The algorithm is inspired by the phase quantification for powder diffraction data using Rietveld refinement (Rietveld, 1967, 1969; Bish, 1988; Kaduk *et al.*, 2021). In this method, powder diffraction patterns are calculated based on known crystal structure solutions while varying the relative quantity of the involved phases. By iteratively fitting theoretical to experimental data, individual relative quantities and peak shapes are refined until calculated and measured diffraction patterns overlap, ultimately leading to an accurate phase quantification. Due to its rapid applicability and automation, the method is widely applied in both research and industry (Degen *et al.*, 2014; Rodríguez-Carvajal,

1993). This study presents a first step in this direction for the texture and phase quantification of thin film samples using GIXD. The key steps involve the extraction of radial line profiles from measured GIXD data combined with the application of intensity correction factors (Gasser *et al.*, 2024). Subsequently, radial line profiles are computed and refined based on known crystal structure solutions. This holistic approach gives accurate phase quantifications combined with detailed information about orientation distributions in a single process.

**2. Thin film pole figures**

To emphasize the meaning of the information contained in a radial line profile, it is helpful to briefly outline the fundamental principles of pole figures. A pole figure represents the orientation distribution of a defined crystallographic lattice plane (Heffelfinger & Burton, 1960) and is frequently presented as a two-dimensional stereogram (Birkholz, 2005). The stereogram is obtained by stereographic projection from a spherical surface, where each point on the surface corresponds to a particular pole direction. Examples of a powder and a sample with uniplanar texture are shown in Fig. 1a and 1b, with the corresponding crystallite orientation spheres given in Fig. 1c and 1d, respectively. A powder is defined by crystallites without preferential orientation, i.e. where each crystallite orientation is present with equal probability, as drawn schematically in Fig. 1a. Consequently, constant pole density is observed over the entire orientation sphere shown in Fig. 1c.

When preparing crystalline thin films, the interaction between the substrate and the crystallites frequently results in a preferential crystallite orientation with respect to the substrate surface (Abdelsamie *et al.*, 2020) as shown schematically in Fig. 1b. This type of texture, with a defined orientation of a crystallographic lattice plane parallel to the substrate, but without in-plane orientation, is called uniplanar texture (the terms fiber texture (Roe & Krigbaum, 1964) or 2D powder (Fischer *et al.*, 2023) are also used in literature). The crystallographic lattice plane parallel to the substrate is typically called the contact plane (Simbrunner *et al.*, 2018). The orientation sphere of a uniplanar textured sample, as shown in Fig. 1d, features concentric rings with constant pole density, caused by the in-plane isotropy of the crystallites. The width of the observed ring is caused by slightly misoriented crystallites and is described by the term out-of-plane mosaicity. A different representation of the sphere of crystallite orientations is given by radial line profiles. Here, the pole density along the orientation sphere is plotted as a function of the polar angle $\psi$. Radial line profiles of a powder sample and a uniplanar textured sample are shown in Fig. 1e and 1f, respectively.

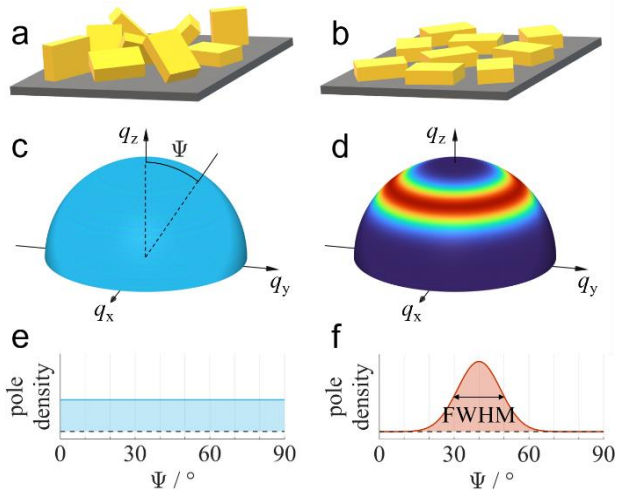

**Figure 1** Schematic presentation of cuboid-shaped crystallites on a substrate with (a) no preferential orientation and (b) uniplanar texture with a defined preferential orientation of a crystallographic lattice plane parallel to the substrate. The sphere of crystallite orientations shows (c) constant pole density for a powder and (d) concentric rings of enhanced pole density for a uniplanar texture, corresponding to an uneven distribution of crystallite orientations. The same behavior is observed in radial intensity profiles showing the pole density as a function of the polar angle $\psi$ for (e) powder and (f) uniplanar texture. The indicated full width at half maximum (FWHM) is used to describe the out-of-plane mosaicity of a uniplanar texture.

Conventionally, pole figures are measured in reflection mode using a point detector. By rotating the sample around two different axes, each point on the orientation sphere is accessible and a complete pole figure can be obtained (Schulz, 1949). For thin films with limited diffraction signal, it is advantageous to measure pole figures in a GIXD geometry (Baker *et al.*, 2010; Schrode *et al.*, 2019). A schematic illustration of a common GIXD setup is shown in Fig. 2a. The sample is mounted on a goniometer that allows rotation around its surface normal and sample tilting to change the angle of incidence $\alpha$ of the X-ray beam on the sample. Diffraction signals are recorded on an area detector. The center of the goniometer is defined as the origin of the detector reference frame $(x, y, z)$.

To illustrate how pole figures are extracted from GIXD measurements, it is helpful to construct the Ewald sphere together with the sphere of crystallite orientations. The radius of the Ewald sphere $\frac{2\pi}{\lambda}$ is fixed by the wavelength $\lambda$ of the X-ray beam, while the radius of the orientation sphere is defined by $q_{hkl}$ of a Bragg peak of interest with Laue indices $hkl$. The center of the orientation sphere is positioned on the surface of the Ewald sphere. The same point defines the origin of the reciprocal sample reference frame $(q_x, q_y, q_z)$. For measurements performed under small angles of incidence, close to the critical angle of total external reflection, the reciprocal sample reference frame and detector reference frame almost overlap, as shown in Fig. 2b. In a single X-ray diffraction measurement, only information along the intersection of the sphere of crystallite orientations with the

Ewald sphere can be accessed. Therefore, to measure a full pole figure, the crystallite orientation sphere needs to be rotated through the Ewald sphere while collecting multiple GIXD patterns. By combining the information contained in the individual measurements, a pole figure can be constructed (Garbe, 2009; Schrode *et al.*, 2019). Additionally, Fig. 2b shows that no features along and near $q_z$ (i.e., perpendicular to the sample surface) are measured in a single GIXD pattern. The minimum accessible polar angle is geometrically given by $\psi_{\min} = (\theta_{hkl} - \alpha)$, where $\theta_{hkl} = \mathrm{asin}\left(\frac{\lambda q_{hkl}}{4\pi}\right)$ is the Bragg angle of the Bragg peak of interest. The missing information can be accessed through a measurement under local specular conditions, where the sample is tilted so that the incidence angle $\alpha$ is equal to the Bragg angle $\theta_{hkl}$ (Baker *et al.*, 2010; Jimison, 2011).

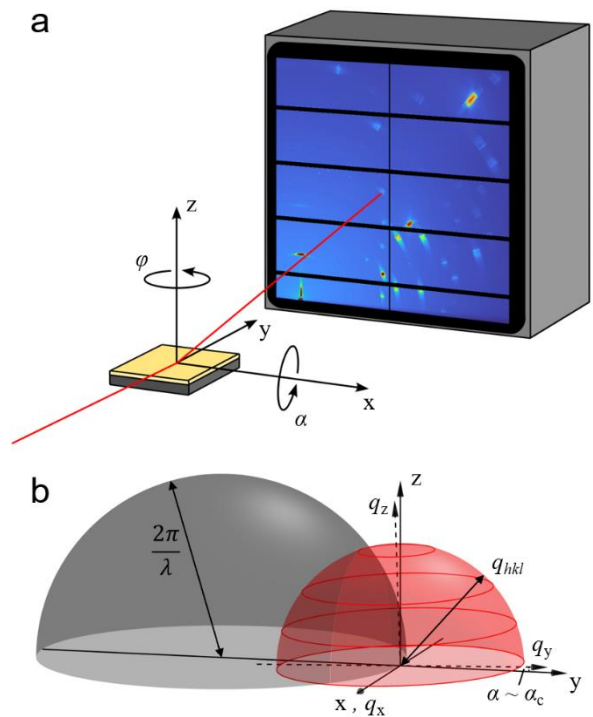

**Figure 2** (a) Scattering geometry for GIXD measurements with an area detector. The incident X-ray beam is fixed and follows the axis y. The sample is mounted on a goniometer and can be rotated around its surface normal through $\varphi$. The angle of incidence $\alpha$ of the primary X-ray beam on the sample is adjusted by tilting the sample around x. (b) Intersection of the Ewald sphere (grey) with the crystallite orientation sphere (schematically shown in red) under grazing incidence conditions.

This work focuses on powders and uniplanar textured samples, as they are most commonly found in thin films. Here, a radial line profile containing all required texture information can be extracted from a single GIXD measurement. For samples with other textures, in-plane isotropy can be artificially created by integrating the detected signal while the sample is rotated around its surface normal (Schrode *et al.*, 2019). Under these conditions, it is useful to represent GIXD patterns in cylindrical reciprocal space coordinates. Due to the in-plane isotropy, the measured intensity can be assumed to be independent of the azimuth angle, allowing a two-dimensional visualization of the measured data in terms of $q_{xy} = \sqrt{q_x^2 + q_y^2}$ and $q_z$ as shown in Fig. 3a.

## 3. Methodology

### 3.1. Radial line profile extraction

Since a quantitative evaluation of diffraction data relies on accurate intensities, a series of intensity correction factors need to be applied to the measured GIXD raw data. These corrections account for the polarization of the incident X-ray beam, the solid angle subtended by each detector pixel, the air absorption between the sample and the detector, and the detector efficiency (Gasser *et al.*, 2024). In addition, a flat-field correction is applied for measurements with a low diffraction signal to remove minor detector artifacts (Jiang, 2015; Schrode *et al.*, 2019). Subsequently, the intensity-corrected two-dimensional GIXD data is transformed into spherical reciprocal space coordinates $I(q, \psi)$ as shown in Fig. 3b. Here, intensity is visualized in terms of the radial component $q = \sqrt{q_{xy}^2 + q_z^2}$ and the polar angle $\psi = \mathrm{atan}\frac{q_{xy}}{q_z}$. In this way, radial line profiles can be extracted along straight lines, simplifying the numerical processing of the data. Transformations between different reciprocal space coordinates are easily accomplished using software such as GIXSGUI (Jiang, 2015), GIDVis (Schrode *et al.*, 2019), PyFAI (Ashiotis *et al.*, 2015), or INSIGHT (Reus *et al.*, 2024).

Before extracting radial line profiles, the background intensity caused by substrate scattering, air scattering, X-ray fluorescence and other effects needs to be subtracted. By linearly interpolating the data above and below the region of interest shown in Fig. 3b, the background intensity can be locally approximated and subtracted from the measured data (Reus *et al.*, 2024). Finally, one-dimensional radial line profiles are obtained by numerical integration of the local background corrected data $I_{bc}(q, \psi)$ between the integration limits $q_i^{min}$ and $q_i^{max}$:

$$I_i^{obs}(\psi) = \int_{q_i^{min}}^{q_i^{max}} I_{bc}(q, \psi)\, q^2 dq \qquad (1)$$

The integration limits $q_i^{min}$ and $q_i^{max}$, as indicated by the white, red, and orange lines in Fig. 3b, must be chosen sufficiently wide to fully include all the peaks involved in the radial line profile. The $q^2$-factor used in the integral arises from the Lorentz correction (Laue, 1926; Buerger, 1940). The Lorentz correction is defined as the inverse of the Jacobian for the transformation from reciprocal cartesian coordinates $q_x$, $q_y$, $q_z$ to the coordinate system in which integration is performed. In the present case, spherical reciprocal space coordinates are used, resulting in the Lorentz correction (Gasser *et al.*, 2024):

$$L = \frac{1}{q^2 \sin \psi} \tag{2}$$

While the $q^2$ of the Lorentz correction is used in eq. 1, the $\sin \psi$ part is included in the fitting algorithm explained below. Its importance for radial line profiles and a phenomenological description are given in the discussion section.

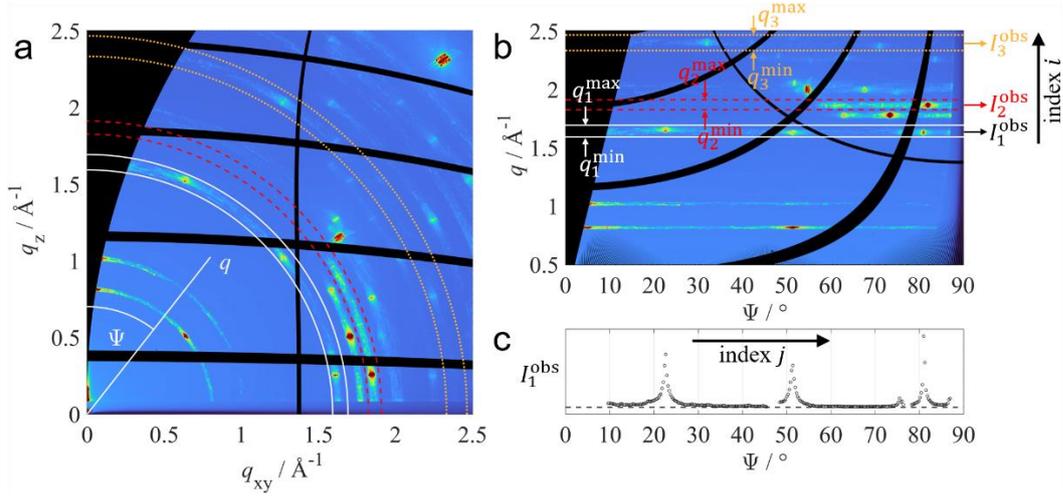

**Figure 3** Measured GIXD data of an anthraquinone thin film presented in (a) cylindrical reciprocal space coordinates $(q_{xy}, q_z)$ and (b) spherical reciprocal space coordinates $(\psi, q)$. The white, red, and orange lines indicate the integration limits $q_i^{min}$ and $q_i^{max}$ used to compute the radial line profiles $I_i^{obs}(\psi)$. Data above and below the limits is used to perform a local background correction of each radial line profile. (c) Radial line profile $I_1^{obs}(\psi)$ obtained after background correction and numerical integration between $q_1^{min}$ and $q_1^{max}$. The individual data points along a radial line profile are indexed with $j$.

### 3.2. Fitting algorithm for quantitative orientation and phase fraction analysis

Quantitative evaluation of measured radial line profiles is achieved by computing and refining theoretical radial line profiles based on a single or multiple known crystal structures. Correspondingly, for every extracted radial line profile $I_i^{obs}(\psi)$, a calculated radial line profile $I_i^{calc}(\psi)$ needs to be computed. For a sample consisting of a single phase with a texture that is either a perfect powder or purely uniplanar with a single contact plane, the calculated intensity along a radial line profile can be expressed as:

$$I_i^{calc}(\psi) = K \sum_{hkl} I_{hkl} \cdot Y(\psi, \sigma, \psi_{hkl}) \tag{3}$$

Here, $K$ is a scale factor and $Y(\psi, \sigma, \psi_{hkl})$ is the intensity distribution function for a Bragg peak with Laue indices $hkl$. The Bragg peak is located at the reciprocal spherical coordinates $q_{hkl}$ and $\psi_{hkl}$ and has an integrated intensity $I_{hkl}$. Details on the calculation of Bragg peak positions $(q_{hkl}, \psi_{hkl})$ from a known crystal structure are given in the Supporting Information S1. In eq. 3, the summation over $hkl$ includes all the Bragg peaks which have their respective $q_{hkl}$ within the integration limits $q_i^{min}$ and $q_i^{max}$ used for the calculation of $I_i^{obs}(\psi)$ in eq. 1. The integrated intensity $I_{hkl}$ of a Bragg peak with Laue indices $hkl$ is calculated via (Warren, 1990):

$$I_{hkl} = I_0 r_e^2 \lambda^3 \frac{V}{V_u^2} \cdot m_{hkl} \cdot |F_{hkl}|^2 = \frac{K}{V_u^2} \cdot m_{hkl} \cdot |F_{hkl}|^2 \tag{4}$$

Here, $I_0$ is the intensity of the primary X-ray beam, $r_e$ is the classical electron radius, $\lambda$ is the X-ray wavelength, $V$ is the diffracting crystalline volume, $V_u$ is the volume of the unit cell, $m_{hkl}$ is the peak multiplicity factor, and $F_{hkl}$ is the structure factor. Correction factors like polarization correction or sample-pixel distance correction were already applied to the measured data and are therefore not considered here. It is important to note that the peak multiplicity is different for powders and uniplanar textured films, as described in more detail in the Supporting Information S1. Some of the given quantities, i.e. $I_0$, $r_e$, $\lambda$, and $V$, can be treated as constants throughout a measurement and are therefore included in the scale factor $K$. The volume of the unit cell $V_u$ needs to be used explicitly for the calculation of the intensity of a Bragg peak, as it varies for different crystalline phases that could be present within the same thin film.

The intensity distribution function $Y(\psi, \sigma, \psi_{hkl})$ in eq. 3 can be any arbitrary function that fits the measured data. For a material with uniplanar texture, it generally follows the shape of a peak function $P(\psi, \sigma, \psi_{hkl})$:

$$Y(\psi, \sigma, \psi_{hkl}) = \frac{1}{N} P(\psi, \sigma, \psi_{hkl}) \tag{5}$$

Here, $\psi_{hkl}$ corresponds to the calculated peak position and $\sigma$ describes the width of the peak and consequently, the out-of-plane mosaicity of crystallites with a certain contact plane. Prominent peak shape functions are Gaussian, Lorentzian, Pseudo-Voigt (Wertheim *et al.*, 1974) or Pearson VII (Hall *et al.*, 1977) functions. Mathematical expressions are provided in the Supporting Information S2. For a powder sample without preferential orientation, the intensity distribution function can be assumed to be constant:

$$Y(\psi, \sigma, \psi_{hkl}) = \frac{1}{N} \tag{6}$$

For both cases, the normalization constant $N$ is implicitly defined through the integral

$$\int_0^\pi Y(\psi, \sigma, \psi_{hkl}) \sin \psi \, d\psi = 1 \tag{7}$$

where the $\sin \psi$ factor is part of the Lorentz correction discussed above. The peak shape functions given in eq. 5 and 6 can be used to compute radial line profiles as shown in Fig. 1e and 1f, respectively.

In general, a thin film can contain multiple phases which are measured together in a GIXD pattern. Similarly, thin film textures are frequently complex and consist of multiple components, i.e. a mixture of powder and uniplanar texture, or multiple uniplanar textures with different contact planes. In such cases, eq. 3 is expanded to:

$$I_i^{\text{calc}}(\psi) = \sum_p K_p \sum_{hkl} I_{p,hkl} \cdot Y_p(\psi, \sigma_p, \psi_{p,hkl}) \tag{8}$$

Here, a summation over $p$ is introduced to include Bragg peaks attributed to different phases or different texture components. Consequently, $K_p$ is the scale factor of a single phase with a single texture component that is described by the intensity distribution functions $Y_p(\psi, \sigma_p, \psi_{p,hkl})$ with integrated intensities $I_{p,hkl}$. Following eq. 4 and 8 it can be concluded that the scale factor $K_p$ is directly proportional to the volume of diffracting material $V_p$ of a component $p$:

$$K_p \propto V_p \tag{9}$$

Since each phase and each texture component is included in the fitting algorithm with its own scale factor $K_p$, accurate relative volume fractions of different phases as well as different texture components can be directly obtained by comparing the scale factors $K_p$ of each component.

Additionally, summing up all scale factors $K_p$ obtained from a single measurement gives a quantity proportional to the total volume of diffracting material on the film:

$$V_{\text{tot}} \propto \sum_p K_p \tag{10}$$

The relative total volume $v_{\text{tot}}$ of diffracting material is subsequently determined by dividing the total crystalline volumes of two samples. This provides a useful quantity for the comparison of differently processed films containing the same crystalline phases and follows a similar definition as the relative degree of crystallinity frequently reported in literature (Mahmood & Wang, 2020; Fischer *et al.*, 2023).

In order to determine the scale factors $K_p$, and correspondingly the volume fractions of the involved phases and texture components, the difference between measured radial line profiles $I_i^{\text{obs}}(\psi)$ and calculated radial line profiles $I_i^{\text{calc}}(\psi)$ needs to be minimized. The minimization function we chose is mathematically expressed following a least squares fitting:

$$\Omega = \sum_{i,j} \left(\frac{1}{q_i^2}|I_i^{\text{obs}}(\psi_j) - I_i^{\text{calc}}(\psi_j)|\right)^2 \tag{11}$$

Here, the index $i$ refers to the different radial line profile extracted from the same GIXD pattern and the index $j$ to the individual datapoints along a radial line profile as shown in Fig. 3c. The weighting factor $\frac{1}{q_i^2}$, defined as the average of $q_i^{\min}$ and $q_i^{\max}$, is used to restore original weights after the multiplication of the measured data with $q^2$ in eq. 1.

There exist several well-established numerical methods to perform the minimization of the function $\Omega$. For the present work, a simulated annealing approach (Kirkpatrick *et al.*, 1983; Černý, 1985; Bertsimas & Tsitsiklis, 1993; Rutenbar, 1989) using stochastic sampling was applied. Simulated annealing is an efficient numerical tool to approximate the global minimum of a function within relatively short timeframes. This makes it an ideal candidate for applications where large numerical data sets require fast and reliable evaluation.

## 4. Experimental results

In the following a detailed description of different applications of the fitting algorithm for quantitative orientation and phase fraction analysis will be given. The three systematically chosen examples include anthraquinone films of a single crystalline phase with different contact planes, zeolitic imidazolate framework-8 (ZIF-8) thin films consisting of both oriented and unoriented crystallites, and binaphthalene films containing two phases with different textures. Experimental details on the GIXD measurements and sample preparation are given in the Supporting Information S3 and S4.1-S6.1.

### 4.1. Orientation quantification of anthraquinone thin films

Anthraquinone thin films were prepared by dip coating using two different withdrawal velocities, 1 μm/s and 2 μm/s. To estimate the involved contact planes, GIXD measurements were performed resulting in the diffraction patterns shown in Fig. 4a and 4e for the samples prepared at 1 μm/s and 2 μm/s, respectively. From the peak positions $\psi_{hkl}$ observed in the diffraction data, three contact planes (001), (100) and (10-2) were identified. 12 radial line profiles were extracted for quantitative analysis of both measurements shown in the Supporting Information S4.2 together with the parameters obtained after fitting. A selection of three radial line profiles extracted between the integration limits shown in Fig. 3 are presented in Fig. 4b-d and 4f-h for the two samples, respectively.

To fit the measured data, peak positions and peak intensities were calculated from the known crystal structure solution (Lonsdale *et al.*, 1966) assuming three contact planes (001), (100) and (10-2). For each component, peak shapes were assumed to follow Pearson VII functions. Peak widths ($\sigma$ and $\eta$ as introduced in eq. 15 of the Supporting Information S2) were assumed to be equal for each contact plane. The calculated radial line profiles, shown in red in Fig. 4b-d and 4f-h, are in good agreement with the measured data. From the obtained scale factors, relative volume fractions of crystallites with the given contact planes were calculated. For the thin film grown at 1 μm/s it was determined that 82% of the crystallites have a (10-2) contact plane, 11% a (001) contact plane and 7% a (100) contact plane. In contrast, the orientation distribution of the sample prepared at 2 μm/s is significantly different, with only 11% of crystallites having a (10-2) contact plane, but 54% having a (001) contact plane and 35% having a (100) contact plane. The out-of-plane mosaicity of the crystallites is small in both cases, giving peak shapes with a FWHM of 0.45° for the 1 μm/s sample and 0.46° for the 2 μm/s sample. The total crystalline volume of the sample prepared at the higher withdrawal velocity of 2 μm/s is significantly lower (nominally 43%) compared to the sample prepared at 1 μm/s, indicating a reduced film thickness. Although it is generally assumed that the film thickness in dip coating increases at higher withdrawal velocities, this tendency is frequently reversed at ultra-low withdrawal velocities as used in the shown example (Scriven, 1988; Grosso, 2011).

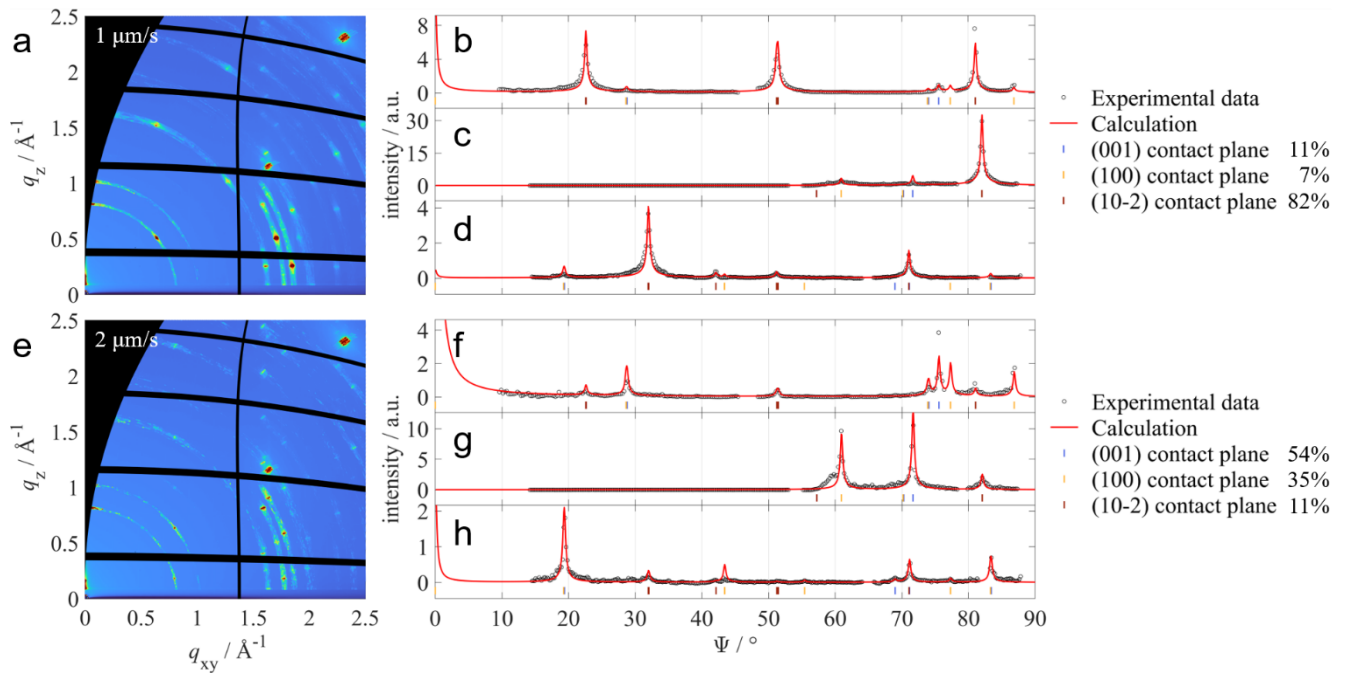

**Figure 4** Measured GIXD patterns of anthraquinone thin films prepared by dip coating with varying withdrawal velocities of (a) 1 μm/s and (e) 2 μm/s. From the measured GIXD data, radial line profiles were extracted at (b), (f) $q \in [1.587, 1.697]$ Å$^{-1}$, (c), (g) $q \in [1.830, 1.910]$ Å$^{-1}$ and (d), (h) $q \in [2.340, 2.490]$ Å$^{-1}$. Calculated line profiles are shown as red lines. The results of the quantitative texture analysis assuming three contact planes are given in the respective legends. Theoretical Bragg peak positions corresponding to the different contact planes are indicated by colored lines below the radial line profiles.

## 4.2. Orientation quantification of ZIF-8 thin films

ZIF-8 is a metal-organic framework (MOF) consisting of zinc nodes connected by 2-methylimidazolate molecules (Park *et al.*, 2006). Fig. 5a shows the GIXD pattern of a ZIF-8 thin film prepared by molecular layer deposition on a bare silicon substrate (Smets *et al.*, 2023). It features Debye-Scherrer rings with enhanced intensities in certain regions along each ring, suggesting a texture composed of an unoriented powder-like fraction and an oriented uniplanar fraction with a (001) contact plane. For quantitative analysis of the two components, a radial line profile was extracted between the integration limits shown in the figure. The measured data of the radial line profile is presented together with the fitted line profile in Fig. 5b. Peak positions and intensities were calculated from the known crystal structure solution of ZIF-8 (Park *et al.*, 2006). For the oriented fraction, Gaussian-shaped peaks were assumed. Relative volume fractions of oriented and unoriented crystallites were calculated from the scale factors, yielding 46% of preferentially oriented material. With respect to the out-of-plane mosaicity of the oriented component, a FWHM of 30.1° was determined.

Fig. 5c shows the GIXD pattern of a ZIF-8 thin film prepared using the same deposition parameters on a gold-covered silicon substrate functionalized with a self-assembled monolayer of octadodecanethiol (ODT) molecules (Smets *et al.*, 2025). Here, significantly sharper peaks can be observed compared to the sample prepared on bare silicon. The increased intensity along the $q_z$-axis is caused by the reflectivity signal of the gold layer on the substrate and can be ignored for this study. However, data at small values of $\psi < 15°$ had to be removed from the extracted radial line profile in Fig. 5d due to an unsuccessful background correction in this regime. By fitting the extracted radial line profile, a high degree of preferential orientation of 87% was obtained, combined with a significantly reduced mosaicity with a FWHM of only 16.5°.

For a better statistical overview of the sample, additional radial line profiles were extracted from each GIXD pattern shown in the Supporting Information S5.2. Fitting revealed similar results, giving 56% and 50% of the oriented fraction for the sample prepared on bare silicon and 93% and 92% of the oriented component for the sample prepared on the ODT-functionalized substrate. Similarly, the FWHM values show little difference with respect to the results presented, giving 29.8° and 28.3° for the silicon substrates compared to 17.0° and 16.2° for the ODT-functionalized substrates.

In contrast to other literature about orientation quantifications (Fischer *et al.*, 2023; Reus *et al.*, 2022), the areas of the differently colored fractions in Fig. 5 are not directly proportional to the corresponding relative volumes. In order to achieve such a correlation, both measured and computed radial line profiles need to be multiplied with a factor $\sin \psi$ (Hammond *et al.*, 2011; Reus *et al.*, 2024). Radial line profiles, where the areas of the different components are proportional to the respective volume fractions, are provided in the Supporting Information S5.2.

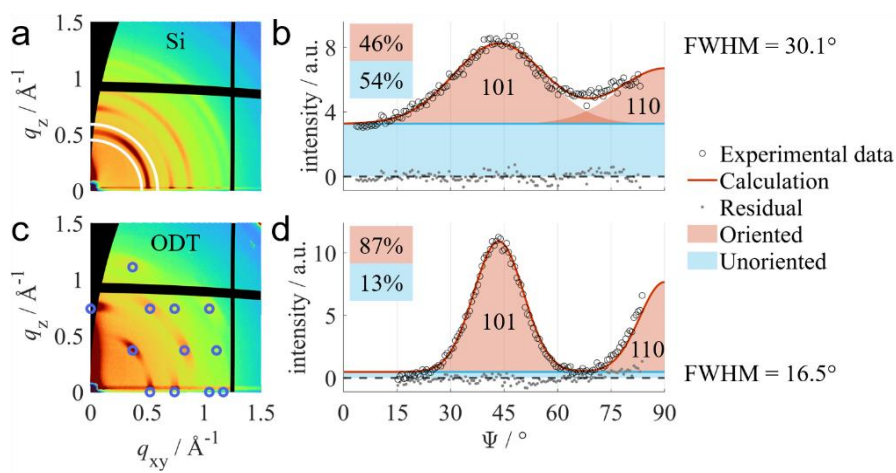

**Figure 5** Measured GIXD patterns of ZIF-8 thin films prepared by molecular layer deposition on (a) a bare silicon substrate and (c) a substrate functionalized with a self-assembled monolayer of ODT molecules. The solid white lines indicate the integration limits for the extraction of radial line profiles. The blue circles correspond to calculated peak positions assuming a (001) contact plane. Extracted radial line profiles for the sample prepared on (b) bare silicon and (d) an ODT-functionalized substrate are fitted assuming a thin film consisting of unoriented (blue) and preferentially oriented crystallites (red). Bragg peaks of the oriented fraction are indexed with their Laue indices.

### 4.3. Phase fraction and orientation quantification of binaphthalene thin films

1,1'-binaphthalene is a model molecule to study axial chirality. The chirality is attributed to a restricted rotation around the bond connecting the naphthalene groups, resulting in two enantiomers with opposite dihedral angles (Pu, 1998). When crystallized, two stable forms of binaphthalene are reported: a racemate consisting of equal amounts of both enantiomers and a chiral form containing a single enantiomer only (Kress *et al.*, 1980).

When preparing binaphthalene thin films by spin coating, it was observed that both chiral and racemic phases crystallize individually on the surface. Fig. 6abc shows a selection of corresponding GIXD patterns of binaphthalene thin films prepared by spin coating at varying spin velocities of 500 rpm, 1000 rpm and 6000 rpm. Here, the white circles correspond to calculated peak positions of the racemic phase with (100) contact plane and the red circles refer to the chiral phase with (127) contact plane. Each measurement shows the presence of both phases, however with a tendency of decreasing intensity of Bragg peaks related to the racemic phase when increasing the spin velocity. Further GIXD patterns for samples prepared at 2000 rpm and 4000 rpm spin velocity are presented in the Supporting Information S6.2.

For quantitative analysis, radial line profiles were extracted from each measurement at two positions indicated by the white and red annular regions in Fig. 6b. The obtained radial line profiles for the sample prepared at 1000 rpm spin velocity are shown in Fig. 6d for the white annular region corresponding to a peak of the racemic phase and Fig. 6e for the red annular region related to peaks of the chiral phase. The extracted radial line profiles for all measurements are shown in the Supporting Information S6.2, together with a table including all the obtained fit parameters. For the fitting algorithm, a uniplanar racemic phase with (100) contact plane and Pseudo-Voigt peaks was assumed. For the chiral phase, a texture consisting of an unoriented powder-like fraction and an oriented uniplanar fraction with (127) contact plane and Gaussian peaks gave the best fitting results. The calculated radial line profiles in Fig. 6de show good agreement with the measured data, giving 33% of racemic binaphthalene, 21% of the oriented and 46% of unoriented chiral binaphthalene for the sample prepared at 1000 rpm spin velocity.

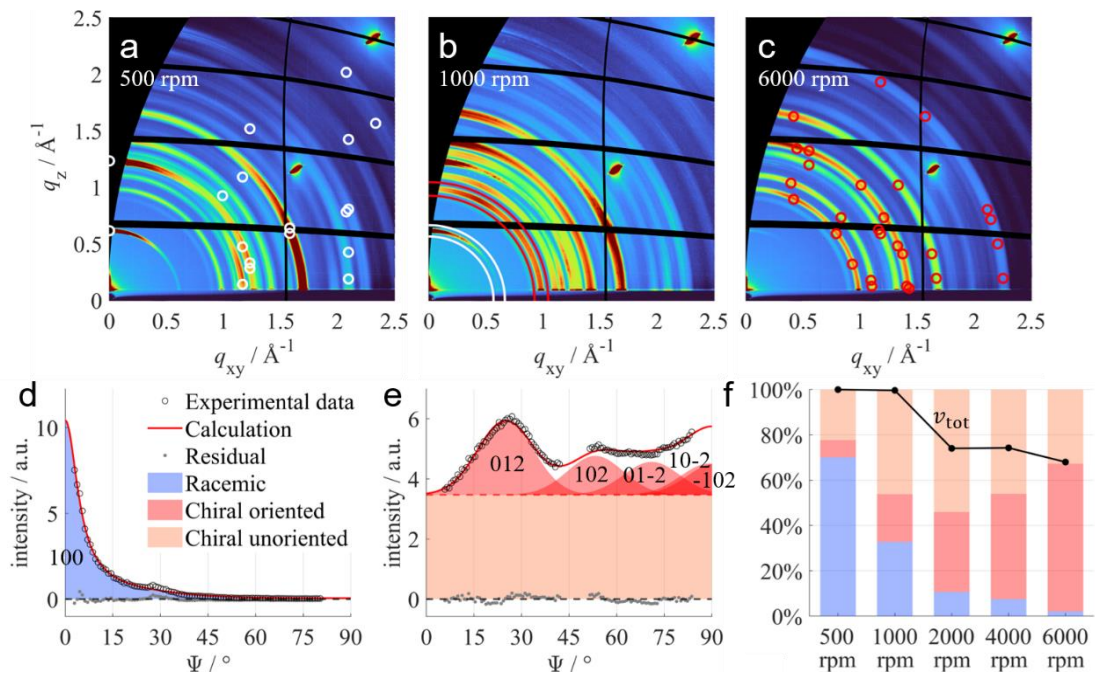

**Figure 6** GIXD patterns of binaphthalene thin films prepared via spin coating at varying spin velocities of (a) 500 rpm, (b) 1000 rpm and (c) 6000 rpm. Calculated peak positions are visualized as white circles for the racemic phase with (100) contact plane and as red circles for the chiral phase with (127) contact plane. Radial line profiles of the 1000 rpm sample were extracted from the (d) white annular region for the racemic phase and (e) red annular region for the chiral phase. The extracted radial line profiles are fitted assuming a racemic phase with (100) contact plane (blue) and a chiral phase consisting of unoriented (orange) and oriented crystallites with a (127) contact plane (red). Bragg peaks of both oriented components are indexed with their Laue indices. (f) The results of the quantitative analysis for the entire series are summarized in a bar chart showing the relative volume fractions of the involved components in blue, red and orange. Additionally, for each sample the relative total volume $v_{tot}$ related to the sample prepared at 500 rpm spin velocity is presented in black.

A summary of the quantitative analysis of the binaphthalene thin films prepared at varying spin velocities is shown in Fig. 6f. A strong decrease of the fraction of the racemic phase is observed with increasing spin velocity, ranging from 70% at 500 rpm to only 1% at 6000 rpm. The simultaneous increase of the chiral phase is mainly attributed to its oriented component, which increases from 8% at 500 rpm to 66% at 6000 rpm. In addition to the volume fractions of the involved components, the total volume was calculated for each sample and compared with the sample prepared at 500 rpm spin velocity. The obtained relative total volumes $v_{\text{tot}}$, shown in black in Fig. 6f, follow a clear trend of decreasing volume and hence, film thickness with increasing spin velocity. However, the 32% reduction in thin film thickness when increasing the spin velocity from 500 rpm to 6000 rpm is significantly less compared to literature models which suggest a 71% reduction for the same parameters (Meyerhofer, 1978; Daughton & Givens, 1982).

**5. Discussion**

The described examples provide a comprehensive and systematic overview of various systems in which a quantitative analysis of radial line profiles can be of interest. In the example of anthraquinone, three distinct preferential orientations were evaluated through the simultaneous fitting of multiple radial line profiles. This way, the impact of data outliers could be minimized and an averaged result over the entire GIXD map was achieved. Furthermore, a significant advantage of using multiple radial line profiles is that missing information or peak overlaps in one radial line profile are compensated with information from other radial line profiles.

For ZIF-8 the relative volume fractions of unoriented and preferentially oriented crystallites were determined. In this case, individual fits were performed on multiple radial line profiles. The quantitative information obtained from the individual fits show good agreement, with deviations below 5% relative volume fractions of the components involved. This variation also gives a rough error estimate, which is mainly attributed to the background correction performed when extracting a radial line profile. Furthermore, the out-of-plane mosaicity of the oriented fraction was calculated, giving a useful additional parameter to describe the quality of an obtained thin film.

The example of binaphthalene shows a combined quantification of both textures and phase fractions present on a thin film. In literature, phase quantifications on GIXD measurements are typically performed by extracting a powder pattern $I(q)$ through integration over the polar angle $\psi$ (Reus *et al.*, 2022). Although such analysis is useful and builds up on the well-established Rietveld refinement method, it is restricted to phases without or only weak preferential orientation. In the case of well preferentially oriented phases, as frequently present on thin films, the missing data above the missing wedge or below the materials horizon might lead to wrong conclusions. For such samples, fitting radial line profiles as described in this work allows to overcome these limitations, presumably leading to more accurate results (Fratschko *et al.*, 2024).

The calculated total volumes $v_{tot}$ give a good and quick estimate of the amount of diffracting material present on a thin film. However, it should be noted that an accurate comparison of different measurements is challenging for GIXD, because measured intensities are affected by variations in the intensity of the primary beam as well as the angle of incidence and several other GIXD-specific peculiar experimental effects (Savikhin *et al.*, 2020). In particular, the angle of incidence has a direct effect on the beam footprint as well as the penetration depth and hence, the volume of material 'observed' in the diffraction experiment.

For systems giving rise to many diffraction peaks, and for GIXD measurements performed at small angles of incidence, overlaps between different Bragg peaks can make the extraction of radial line profiles with appropriate background correction difficult. In such cases, a method to directly compute and fit GIXD patterns in 2D would be highly desirable. Although some software packages exist to calculate 2D GIXD data (Breiby *et al.*, 2008; Savikhin *et al.*, 2020), they are still at an early stage of development and an automatic fitting of measured data has not yet been achieved.

Given that the Lorentz correction as a Jacobian for integral transformations (i.e. $L^{-1} = q^2 \sin\psi$ for spherical reciprocal space coordinates) is not overly intuitive, it is useful to consider the phenomenological basis for the "$\sin\psi$ correction" in some more detail, particularly because it may seem counterintuitive at first that it is also necessary for powder samples where the corresponding $\sin\psi$-corrected intensity is not uniform along $\psi$, even though the sample is a perfect powder. Phenomenologically, the $\sin\psi$ factor is typically considered as a correction to account for the difference between the total scattered intensity of a material presented in the orientation sphere as shown in Fig. 1cd compared to its corresponding radial line profile shown in Fig. 1ef (Baker *et al.*, 2010; Jimison, 2011). Upon inspection of the Ewald sphere representation in Fig. 2b, it is evident that the radial line profile obtained from a single GIXD experiment provides access to only a small fraction of the orientation sphere and correspondingly represents only a small fraction of the total scattered intensity of the material present on the thin film. A conceptual equivalent argument can also be made in a purely real space representation: For a fully uniplanar thin film, almost all lattice planes parallel to the substrate fulfil the Bragg condition for a given Laue-index. In contrast, most of the lattice planes perpendicular to the substrate do not fulfil the Bragg condition because they do not have the correct azimuthal orientation.

For the phenomenological discussion of the "$\sin\psi$ correction", it is also instructive to recall that it is applied to data in spherical reciprocal space coordinates $I(q,\psi)$ and that the polar angle $\psi$ is defined as $\psi = \mathrm{atan}\frac{q_{xy}}{q_z}$, containing the ratio between in-plane and out-of-plane scattering vector components. This conceptually implies that the respective radial line profile analysis in $(q,\psi)$-space is related to the relative distribution of crystallites along the $q_z$ axis versus the $q_{xy}$ plane. In other words, the orientations are referenced against an axis in one case (the surface normal $q_z$) and against a plane in the other case (the substrate plane $q_{xy}$). This phenomenologically rationalizes the necessity of the "$\sin\psi$ correction" also for a powder. In this argument, the factor $\sin\psi$ is simply the relative intersection of the Ewald sphere with the radius of the circle along the surface of the orientation sphere at a height that belongs to $\psi$ and correspondingly scales with $\sin\psi$ (Hammond *et al.*, 2011).

To conclude the phenomenological discussion of the "$\sin\psi$ correction": Under the assumption of a sample with in-plane isotropy, the total amount of intensity and hence, material can be deduced by multiplying a measured radial line profile with a factor $\sin\psi$ (Fischer *et al.*, 2023; Reus *et al.*, 2024). It must be pointed out however, that the $\sin\psi$ factor is not specific to the quantification of textures. In fact and importantly, the $\sin\psi$ correction is part of the Lorentz correction as described in eq. 2, and is therefore generally essential when determining Bragg peak intensities from GIXD measurements (Gasser *et al.*, 2024). In this work we show that, when fitting radial line profiles with a calculated line profile based on the structure factors of a known crystal structure solution, reliable fits can only be achieved after including the Lorentz correction in eq. 1 and eq. 7. Although the mathematical description of the Lorentz factor as a Jacobian for integral transformations might intuitively seem different to the phenomenological description of the "$\sin\psi$ correction" as a correction for texture quantifications, the provided experimental examples show that both corrections are indeed equivalent.

## 6. Conclusion

This work presents a detailed insight into the quantification of textures and phase fractions through the evaluation of radial line profiles. Three practical applications were shown, including anthraquinone thin films with uniplanar texture but different contact planes, ZIF-8 MOF thin films with an oriented and unoriented component, and binaphthalene thin films containing two phases with different textures. For each system, reliable relative volume fractions of the components involved could be determined, and it was shown how the calculation of additional parameters such as mosaicity and total volume can be useful to compare different thin films containing the same compound. A significant advantage of the presented fitting algorithm is that the fit function takes into account a radial line profile extrapolated beyond experimental limits like the missing wedge and the materials horizon. This way, accurate results can be obtained without having to combine data from measurements at different angles of incidence, which is of particular interest for in situ and operando studies.

The presented algorithm provides a systematic and widely applicable tool for the computing and fitting of radial line profiles. Through the obtained information, a deep insight into the quality and composition of a thin film is achieved in a quantitative way. Further development could result in a promising path for a more streamlined evaluation of experimental GIXD data.

**Acknowledgements**    We acknowledge Elettra Sincrotrone Trieste for providing access to its synchrotron radiation facilities and we thank Luisa Barba and Giorgio Bais for assistance in using beamline XRD1 (proposals 20220410, 20225028 and 20240009). We also acknowledge Jesús Gándara-Loe for supporting the GIXD measurements of ZIF-8.

Author contributions: sample preparation and performance of experiments, FG, SJ and JSm; algorithm development and data evaluation, FG; theoretical considerations, FG, MF and JSi; writing of the manuscript, all authors; supervision of the research, HGS and RR.

**Funding Information**

This research was funded in whole, or in part, by the Austrian Science Fund (FWF) 10.55776/P34463. For the purpose of open access, the author has applied a CC BY public copyright license to any Author Accepted Manuscript version arising from this submission. Furthermore, the research leading to this result has been co-funded by the project NEPHEWS under Grant Agreement No 101131414 from the EU Framework Program for Research and Innovation Horizon Europe. JSm acknowledges the support of FWO Vlaanderen for the fellowship 11H8123N. VRG acknowledges the funding by Generalitat Valenciana through the Plan Gent-T of Excellence (CIDEIG/2022/32). HGS acknowledges the funding by the German Federal Ministry of Education and Research (BMBF) and the Ministry of Economic Affairs, Industry, Climate Action and Energy of the State of North Rhine-Westphalia through the project HC-H2, and from the BMBF via Projects 05K22PP1 and 05K24CJ1.

**References**

Abdelsamie, M., Xu, J., Bruening, K., Tassone, C. J., Steinrück, H.-G. & Toney, M. F. (2020). *Advanced Functional Materials* **30**, 2001752.

Ashiotis, G., Deschildre, A., Nawaz, Z., Wright, J. P., Karkoulis, D., Picca, F. E. & Kieffer, J. (2015). *J Appl Cryst* **48**, 510–519.

Bai, Y., Liu, C., Chen, T., Li, W., Zheng, S., Pi, Y., Luo, Y. & Pang, H. (2021). *Angewandte Chemie International Edition* **60**, 25318–25322.

Baker, J. L., Jimison, L. H., Mannsfeld, S., Volkman, S., Yin, S., Subramanian, V., Salleo, A., Alivisatos, A. P. & Toney, M. F. (2010). *Langmuir* **26**, 9146–9151.

Bertsimas, D. & Tsitsiklis, J. (1993). *Statistical Science* **8**, 10–15.

Birkholz, M. (2005). Thin Film Analysis by X-Ray Scattering Wiley-VCH.

Bish, D. L. (1988). *Journal of Applied Crystallography* **21**, 86–91.

Breiby, D. W., Bunk, O., Andreasen, J. W., Lemke, H. T. & Nielsen, M. M. (2008). *J Appl Cryst* **41**, 262–271.

Buerger, M. J. (1940). *Proceedings of the National Academy of Sciences* **26**, 637–642.

Černý, V. (1985). *J Optim Theory Appl* **45**, 41–51.

Chou, K. W., Yan, B., Li, R., Li, E. Q., Zhao, K., Anjum, D. H., Alvarez, S., Gassaway, R., Biocca, A., Thoroddsen, S. T., Hexemer, A. & Amassian, A. (2013). *Advanced Materials* **25**, 1923–1929.

Daughton, W. J. & Givens, F. L. (1982). *J. Electrochem. Soc.* **129**, 173.

Degen, T., Sadki, M., Bron, E., König, U. & Nénert, G. (2014). *Powder Diffraction* **29**, S13–S18.

Fischer, J. C., Li, C., Hamer, S., Heinke, L., Herges, R., Richards, B. S. & Howard, I. A. (2023). *Advanced Materials Interfaces* **10**, 2202259.

Fratschko, M., Zhao, T., Fischer, J. C., Werzer, O., Gasser, F., Howard, I. A. & Resel, R. (2024). *ACS Appl. Nano Mater.* **7**, 25645–25654.

Garbe, U. (2009). *J Appl Cryst* **42**, 730–733.

Gasser, F., Simbrunner, J., Huck, M., Moser, A., Steinrück, H.-G. & Resel, R. (2024). *J Appl Cryst*.

Greco, A., Hinderhofer, A., Dar, M. I., Arora, N., Hagenlocher, J., Chumakov, A., Grätzel, M. & Schreiber, F. (2018). *J. Phys. Chem. Lett.* **9**, 6750–6754.

Grosso, D. (2011). *J. Mater. Chem.* **21**, 17033–17038.

Grott, S., Kotobi, A., Reb, L. K., Weindl, C. L., Guo, R., Yin, S., Wienhold, K. S., Chen, W., Ameri, T., Schwartzkopf, M., Roth, S. V. & Müller-Buschbaum, P. (2022). *Solar RRL* **6**, 2101084.

Guo, H., Zhu, G., Hewitt, I. J. & Qiu, S. (2009). *J. Am. Chem. Soc.* **131**, 1646–1647.

Hall, M. M., Veeraraghavan, V. G., Rubin, H. & Winchell, P. G. (1977). *J Appl Cryst* **10**, 66–68.

Hammond, M. R., Kline, R. J., Herzing, A. A., Richter, L. J., Germack, D. S., Ro, H.-W., Soles, C. L., Fischer, D. A., Xu, T., Yu, L., Toney, M. F. & DeLongchamp, D. M. (2011). *ACS Nano* **5**, 8248–8257.

Heffelfinger, C. J. & Burton, R. L. (1960). *Journal of Polymer Science* **47**, 289–306.

Jiang, Z. (2015). *J Appl Cryst* **48**, 917–926.

Jimison, L. H. (2011). Understanding Microstructure and Charge Transport in Semicrystalline Polythiophenes. Stanford University.

Kaduk, J. A., Billinge, S. J. L., Dinnebier, R. E., Henderson, N., Madsen, I., Černý, R., Leoni, M., Lutterotti, L., Thakral, S. & Chateigner, D. (2021). *Nat Rev Methods Primers* **1**, 1–22.

Kalyani, N. T., Swart, H. C. & Dhoble, S. J. (2017). Principles and Applications of Organic Light Emitting Diodes (OLEDs) Woodhead Publishing.

Khalil, I. E., Fonseca, J., Reithofer, M. R., Eder, T. & Chin, J. M. (2023). *Coordination Chemistry Reviews* **481**, 215043.

Kirkpatrick, S., Gelatt, C. D. & Vecchi, M. P. (1983). *Science* **220**, 671–680.

Köhler, A. & Bässler, H. (2015). Electronic Processes in Organic Semiconductors: An Introduction John Wiley & Sons.

Kress, R. B., Duesler, E. N., Etter, M. C., Paul, I. C. & Curtin, D. Y. (1980). *J. Am. Chem. Soc.* **102**, 7709–7714.

Laue, M. v (1926). *Zeitschrift für Kristallographie - Crystalline Materials* **64**, 115–142.

Li, Z., Huang, X., Sun, C., Chen, X., Hu, J., Stein, A. & Tang, B. (2017). *J Mater Sci* **52**, 3979–3991.

Liang Tan, W., Cheng, Y.-B. & R. McNeill, C. (2020). *Journal of Materials Chemistry A* **8**, 12790–12798.

Liu, Y., Ng, Z., Khan, E. A., Jeong, H.-K., Ching, C. & Lai, Z. (2009). *Microporous and Mesoporous Materials* **118**, 296–301.

Lonsdale, K., Milledge, J. & El Sayed, K. (1966). *Acta Cryst* **20**, 1–13.

Mahmood, A. & Wang, J.-L. (2020). *Solar RRL* **4**, 2000337.

Meyerhofer, D. (1978). *Journal of Applied Physics* **49**, 3993–3997.

Mingabudinova, L. R., Vinogradov, V. V., Milichko, V. A., Hey-Hawkins, E. & Vinogradov, A. V. (2016). *Chemical Society Reviews* **45**, 5408–5431.

Müller-Buschbaum, P. (2014). *Advanced Materials* **26**, 7692–7709.

Ogle, J., Powell, D., Amerling, E., Smilgies, D.-M. & Whittaker-Brooks, L. (2019). *CrystEngComm* **21**, 5707–5720.

Park, K. S., Ni, Z., Côté, A. P., Choi, J. Y., Huang, R., Uribe-Romo, F. J., Chae, H. K., O'Keeffe, M. & Yaghi, O. M. (2006). *Proceedings of the National Academy of Sciences* **103**, 10186–10191.

Paulsen, B. D., Wu, R., Takacs, C. J., Steinrück, H.-G., Strzalka, J., Zhang, Q., Toney, M. F. & Rivnay, J. (2020). *Advanced Materials* **32**, 2003404.

Pu, L. (1998). *Chem. Rev.* **98**, 2405–2494.

Qin, M., Chan, P. F. & Lu, X. (2021). *Advanced Materials* **33**, 2105290.

Reus, M. A., Reb, L. K., Kosbahn, D. P., Roth, S. V. & Müller-Buschbaum, P. (2024). *J Appl Cryst* **57**, 509–528.

Reus, M. A., Reb, L. K., Weinzierl, A. F., Weindl, C. L., Guo, R., Xiao, T., Schwartzkopf, M., Chumakov, A., Roth, S. V. & Müller-Buschbaum, P. (2022). *Advanced Optical Materials* **10**, 2102722.

Richter, L. J., DeLongchamp, D. M., Bokel, F. A., Engmann, S., Chou, K. W., Amassian, A., Schaible, E. & Hexemer, A. (2015). *Advanced Energy Materials* **5**, 1400975.

Rietveld, H. M. (1967). *Acta Cryst* **22**, 151–152.

Rietveld, H. M. (1969). *J Appl Cryst* **2**, 65–71.

Rivnay, J., Mannsfeld, S. C. B., Miller, C. E., Salleo, A. & Toney, M. F. (2012). *Chem. Rev.* **112**, 5488–5519.

Rodríguez-Carvajal, J. (1993). *Physica B: Condensed Matter* **192**, 55–69.

Roe, R. -J. & Krigbaum, W. R. (1964). *The Journal of Chemical Physics* **40**, 2608–2615.

Rutenbar, R. A. (1989). *IEEE Circuits and Devices Magazine* **5**, 19–26.

Savikhin, V., Steinrück, H.-G., Liang, R.-Z., Collins, B. A., Oosterhout, S. D., Beaujuge, P. M. & Toney, M. F. (2020). *J Appl Cryst* **53**, 1108–1129.

Schlepütz, C. M., Herger, R., Willmott, P. R., Patterson, B. D., Bunk, O., Brönnimann, C., Henrich, B., Hülsen, G. & Eikenberry, E. F. (2005). *Acta Cryst A* **61**, 418–425.

Schrode, B., Pachmajer, S., Dohr, M., Röthel, C., Domke, J., Fritz, T., Resel, R. & Werzer, O. (2019). *J Appl Cryst* **52**, 683–689.

Schulz, L. G. (1949). *Journal of Applied Physics* **20**, 1030–1033.

Scriven, L. E. (1988). *MRS Online Proceedings Library* **121**, 717–729.

Simbrunner, J., Simbrunner, C., Schrode, B., Röthel, C., Bedoya-Martinez, N., Salzmann, I. & Resel, R. (2018). *Acta Cryst A* **74**, 373–387.

Smets, J., Cruz, A. J., Rubio-Giménez, V., Tietze, M. L., Kravchenko, D. E., Arnauts, G., Matavž, A., Wauteraerts, N., Tu, M., Marcoen, K., Imaz, I., Maspoch, D., Korytov, M., Vereecken, P. M., De Feyter, S., Hauffman, T. & Ameloot, R. (2023). *Chem. Mater.* **35**, 1684–1690.

Smets, J., Rubio-Giménez, V., Gándara-Loe, J., Adriaenssens, J., Fratschko, M., Gasser, F., Resel, R., Brady-Boyd, A., Ninakanti, R., De Feyter, S., Armini, S. & Ameloot, R. (2025). *Chem. Mater.* **37**, 400–406.

Steele, J. A., Solano, E., Hardy, D., Dayton, D., Ladd, D., White, K., Chen, P., Hou, J., Huang, H., Saha, R. A., Wang, L., Gao, F., Hofkens, J., Roeffaers, M. B. J., Chernyshov, D. & Toney, M. F. (2023). *Advanced Energy Materials* **13**, 2300760.

Tian, Y.-B., Vankova, N., Weidler, P., Kuc, A., Heine, T., Wöll, C., Gu, Z.-G. & Zhang, J. (2021). *Advanced Science* **8**, 2100548.

Warren, B. E. (1990). X-ray Diffraction New York: Dover Publications.

Wertheim, G. K., Butler, M. A., West, K. W. & Buchanan, D. N. E. (1974). *Review of Scientific Instruments* **45**, 1369–1371.

Werzer, O., Kowarik, S., Gasser, F., Jiang, Z., Strzalka, J., Nicklin, C. & Resel, R. (2024). *Nat Rev Methods Primers* **4**, 1–20.